\documentclass[twocolumn,showpacs,preprintnumbers,amsmath,amssymb]{revtex4}

\usepackage{graphicx}
\usepackage{dcolumn}
\usepackage{bm}

\begin{document}

\title{Nonvanishing spin Hall currents in the presence of magnetic impurities}

\author{Pei Wang, You-Quan Li, and Xuean Zhao}

\affiliation{Zhejiang Institute of Modern Physics and Department of Physics,\\
Zhejiang University, Hangzhou 310027, P. R. China\\}

\date{\today}

\begin{abstract}
The intrinsic spin Hall conductivity in a two dimensional electron
gas with Rashba spin-orbit coupling is evaluated by taking account of
anisotropic coupling between magnetic impurities and itinerant electrons.
In our calculation Kubo's linear response
formalism is employed and the vertex correction is considered. In
the semiclassical limit $\mu \gg 1/\tau$, a non-vanishing
spin Hall conductivity $\sigma^{}_{sH}$ is found to depend on the
momentum relaxation time $\tau$, spin-orbit splitting $\Delta$ and
the anisotropic coefficient of interaction between itinerant
electrons and magnetic impurities. The clean limit of
$\sigma^{}_{sH}$ is in the region of $e/8\pi \sim e/6\pi$, depending
on the anisotropic coefficient.
\end{abstract}


\maketitle

\section{Introduction}

Recently the spin Hall effect (SHE)~\cite{zhang,sinova04,Kato,sinova05,Rashba,Loss05,dasSarma06,SFZhang}
has attracted much attention due to its potential application in spintronics.
In the spin Hall effect, a longitudinal electric field creates a transverse
motion of spins with the spin-up and spin-down carriers moving in
opposite directions, which leads to a transverse spin current
perpendicular to the external electric field.
To understand the spin Hall effect,
one needs to study the intrinsic spin Hall effect
(ISHE) that has been discussed intensively.
Theoretically ISHE may exist
in the p-type semiconductor~\cite{zhang} and two-dimensional
electron gas (2DEG)~\cite{sinova04}.
Sinova et al.~\cite{sinova04} predicted a universal spin Hall conductivity
in clean 2DEG as
\begin{equation*}
\sigma_{sH}=\frac{e}{8\pi}.
\end{equation*}
From then on, many researchers have devoted to the issue of
whether this result can be modified in the presence of impurity.
Some authors found that an arbitrarily small concentration of
impurities would suppress the spin Hall conductivity to zero due
to the vertex
corrections~\cite{mishchenko,Inoue03,Inoue04,raimondi,dimitrova}.
While others~\cite{Loss04,Burkov,nomura} argued that the spin Hall
conductivity was robust in the presence of disorder, falling to
zero only when the lifetime broadening is larger than the spin
orbit splitting of the bands, i.e.,  $1/\tau > \Delta$.
Grimaldi~\cite{grimaldi} considered the situation of sufficiently
low electron density, and found that the vertex corrections no
longer suppressed the spin Hall conductivity when the Fermi energy
is comparable to or smaller than the spin-orbit energy. Other
authors revealed that the vanishing of $\sigma_{sH}$ was a
peculiar feature of the linear Rashba model. Taking into account
of nonlinear momentum dependence of the spin-orbit interaction
$\alpha(p)$~\cite{nomura2} or a nonquadratic band spectrum
$\varepsilon(p)$ \cite{krotkov}, $\sigma_{sH}$ was robust against
impurity scattering.

Very recently, an important step was made by Inoue et al.~ \cite{Inoue06}
who found the spin Hall
conductivity was not zero in the existence of magnetic impurities
in the limit of $\Delta \tau \gg 1$. Physically, the acceleration
of the electrons by the external electric field modifies the
SO-induced pseudomagnetic field such that the spins are tilted out
of the 2DEG plane in directions that are opposite for positive and
negative lateral momentum states. This corresponds to a flow of
$\sigma_z=1/2$ and $\sigma_z=-1/2$ spins in opposite directions
without a corresponding net charge transport.
In the presence of
isotropic impurity scattering, the spin Hall current is proportional
to the time derivative of the spin polarization~\cite{dimitrova}
which vanishes in a stationary state. Whereas, this relation does not
fulfil for magnetic impurities, leaving over an opportunity for nonvanishing
spin Hall conductivity.
As we are aware, the case considered by Inoue has not been developed to
anisotropic model in the semiclassical limit $\mu \gg 1/\tau$.
And Inoue et al. adopted a simplified approximation in
calculating the integral of Green's functions. It is therefore
obliged to study such kind models and investigate spin Hall
conductivity carefully.

In this paper we calculate the spin Hall conductivity for a
two-dimensional magnetically disordered Rashba-electron gas where the
magnetic interaction is anisotropic in the limit of large Fermi
circle $\mu \gg 1/\tau, \Delta$. The XXZ-type interaction between
the magnetic impurity and the electron spin is adopted. Our calculation is
carried out by considering the vertex corrections with the help of
Kubo's  linear response formalism within the self-consistent Born approximation.
The paper is organized as follows: In Sec.~\ref{sec:model},
we introduce the model by taking into account of magnetic
impurities. In Sec.~\ref{sec:spinHall}, the calculation procedure of the spin
Hall conductivity is presented. In Sec.~\ref{sec:discussion}, we give a comparison
between our result and the result in other disordered system and
discuss the possible reasons that cause the vanishing spin Hall
conductivity.

\section{The model with magnetic impurity}\label{sec:model}

\begin{figure}[h]
\vspace{1mm}
\includegraphics[width=0.21\textwidth]{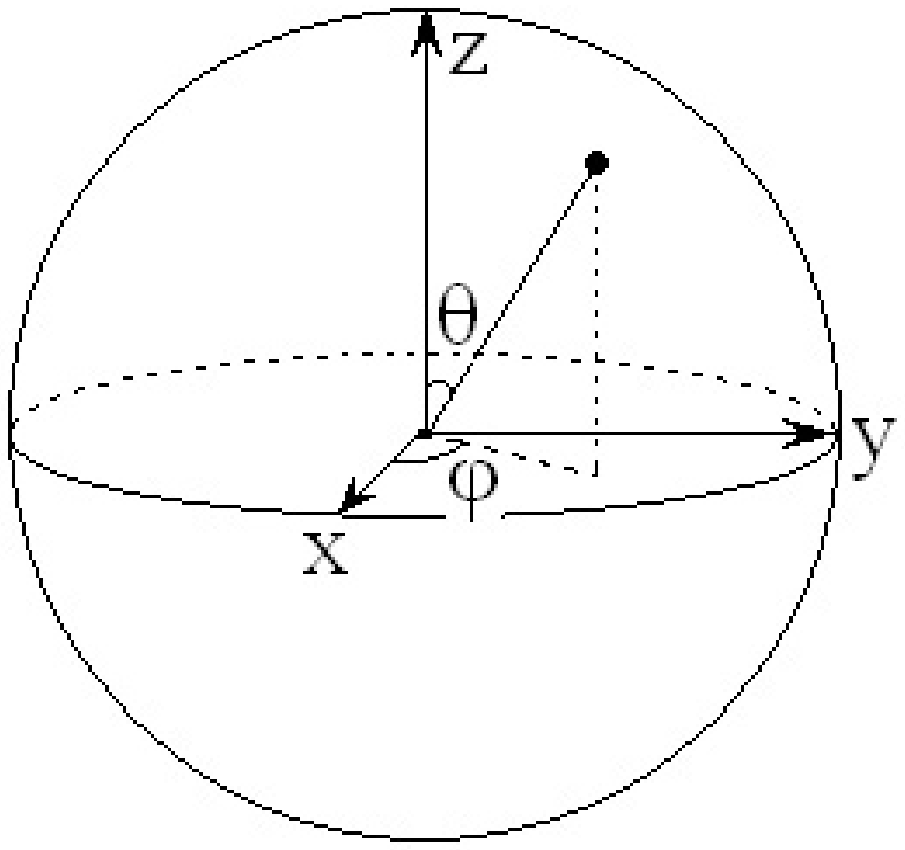}
\caption{\label{fig:spin}Bloch sphere representation of an
magnetic impurity.}
\end{figure}

We consider the 2DEG with Rashba spin-orbit coupling in the
presence of impurities, whose Hamiltonian consists of two parts
$H=H_0+V_{dis}$ with $H_0$ the sum of the kinetic and Rashba
terms, and $V_{dis}$ the potential caused by impurities.
The main contribution of a short-ranged magnetic impurity
can be described by an interaction between an itinerant electron
and a local magnetic moment~\cite{Anderson}
whose orientation is defined by polar
and azimuthal angles $(\theta,\phi)$ (see Fig. \ref{fig:spin}).
A quite general form of such kind interaction is of XXZ type
whose second quantization form is given by the following Hamiltonian
\begin{eqnarray}\label{model}
V_{dis}=&& \sum_{i=1}^N \int d\textbf{r}^2 u \delta(\textbf{r}-\textbf{R}_i) \\
&& \hat{\psi}^\dag(\textbf{r}) \left(
\begin{array}{cc}
  \gamma\cos\theta_i & \sin\theta_i e^{-i\phi_i} \nonumber\\
  \sin\theta_i e^{i\phi_i} & -\gamma\cos\theta_i
\end{array}
\right) \hat{\psi}(\textbf{r}),
\end{eqnarray}
where $(\textbf{R}_i,\theta_i,\phi_i)$ denote the position and
orientation of the $i$th impurity, $N$ the total number of impurities;
$u$ and $\gamma$ refer to, respectively, the strength and the anisotropy
of the coupling between the itinerant electrons and the impurities.
\begin{equation}
\hat{\psi}(r)=\left(
\begin{array}{c}
\hat{\psi}_\uparrow (r) \\
\hat{\psi}_\downarrow (r)
\end{array}\right).
\end{equation}
The interaction becomes isotropic for the special value $\gamma=1$
which corresponds to the model considered by Inoue \cite{Inoue06}.
We assume that the distribution of impurities is homogeneous and
their orientations are isotropic, accordingly,
\begin{eqnarray}\nonumber
&&P(\textbf{R}_1\theta_1\phi_1,\textbf{R}_2\theta_2\phi_2,\cdots,\textbf{R}_N\theta_N\phi_N)
\\&&= (\frac{1}{4\pi V})^N d\textbf{R}_1 d\Omega_1d\textbf{R}_2d\Omega_2\cdots
d\textbf{R}_Nd\Omega_N,
\end{eqnarray}
where $d\Omega_i=\sin\theta_i d\theta_i d\phi_i$
and $V$ denotes the area of the 2DEG.
Note that the model given by Eq.~(\ref{model}) for magnetic impurity
is not reducible to that for nonmagnetic impurity
since the matrix between $\hat{\psi}^\dag(\textbf{r})$ and
$\hat{\psi}(\textbf{r})$ for the former is a traceless matrix while that for the latter
is an unit matrix.
For a general discussion, the strength of Rashba coupling is momentum-dependent
{\it i.e.}, $H_0=\varepsilon(p)+\alpha(p)(\sigma_x p_y-\sigma_y p_x)$.
This Hamiltonian $H_0$ can be diagonalized to be
\begin{eqnarray}
\varepsilon_{\pm}(p)=\varepsilon(p)\mp p\alpha(p),
\end{eqnarray}
by the unitary matrix
\begin{eqnarray}
U(\textbf{p})= \frac{1}{\sqrt{2}}
\left( \begin{array}{cc} 1 & 1 \\
ie^{i\varphi_{\textbf{p} }} & -ie^{i\varphi_{\textbf{p} }}
\end{array} \right),
\end{eqnarray}
where $\varphi_{\textbf{p}}$ denotes the angle between the
momentum $\textbf{p}$ and the $x$ axis.
The momentum-dependent Rashba coupling has important effects on the
spin-Hall conductivity in the presence of nonmagnetic scatterers~\cite{nomura2},
while it is not relevant to the final result in the present case
with magnetic scatterers.

\section{Spin Hall conductivity}\label{sec:spinHall}

As is well known that spin angular momentum is not conserved in
the presence of spin-orbit coupling, and the spin current is
defined as
\begin{eqnarray}\nonumber
J_y^z(\textbf{p})=&&\frac{1}{4}(v_y \sigma_z+\sigma_z v_y)
\\ =&&\frac{1}{2}\frac{d \varepsilon(p)}{dp_y}\sigma_z.
\end{eqnarray}
On the basis of Kubo's formalism, the spin Hall conductivity can
be expressed as~\cite{dimitrova}
\begin{eqnarray}\label{SHE}
\sigma_{sH}(\omega)=
  \displaystyle\frac{e}{\omega V}\int \frac{d\omega_1}{2\pi}\textrm{Tr}
   \Bigl\{n^{}_F(\omega+\omega_1)J_y^z \hspace{21mm}
     \nonumber\\
\bigl[G^r(\omega+\omega_1)- G^a(\omega+\omega_1)\bigr]
    j_x G^a(\omega_1) \hspace{16mm}
      \nonumber\\
 +~n^{}_F(\omega_1) J_y^z G^r(\omega+\omega_1) j_x \bigl[G^r(\omega_1)- G^a(\omega_1)\bigr]
   \Bigr\},\hspace{1mm}
\end{eqnarray}
where $n^{}_F(\omega)=1/(e^{\beta(\omega-\mu)}+1)$ is the Fermi
function, $j_x(\textbf{p})=\displaystyle\frac{d}{d p_x}[\varepsilon(p)+\alpha(p)(\sigma_xp_y-\sigma_yp_x)]$
is the one-particle charge current operator,
$G^r$ and $G^a$ are the retarded and advanced Green's functions.
The trace is taken
over momentum and spin indices. The diagrammatic method is
employed to calculate the average spin Hall conductivity over the
distribution of impurities. The trace in Eq.(\ref{SHE}) is
expanded into the sum of diagrams (see Fig. \ref{fig:SHC}).
\begin{figure*}[ht]
\begin{picture}(0,66)(0,0)
\put(-230,30){$\textrm{Tr}[J_y^z G^r j_x G^a]=$}
\put(-140,30){\begin{picture}(0,0) \put(0,0){\circle*{5}}
\put(60,0){\circle{5}} \qbezier(0,3)(0,25)(30,25)
\qbezier(60,3)(60,25)(30,25) \qbezier(0,-3)(0,-25)(30,-25)
\qbezier(60,-3)(60,-25)(30,-25) \put(30,-35){$\bar{G}^r$}
\put(30,30){$\bar{G}^a$} \put(-15,0){$J^z_y$} \put(65,0){$j_x$}
\end{picture} }
\put(-60,30){$+$} \put(-30,30){\begin{picture}(0,0)
\put(0,0){\circle*{5}} \put(60,0){\circle{5}}
\qbezier(0,3)(0,25)(30,25) \qbezier(60,3)(60,25)(30,25)
\qbezier(0,-3)(0,-25)(30,-25) \qbezier(60,-3)(60,-25)(30,-25)
\put(30,-35){$\bar{G}^r$} \put(30,30){$\bar{G}^a$}
\put(-15,0){$J^z_y$} \put(65,0){$j_x$}
\multiput(30,0)(0,2.5){11}{\circle*{1}}
\multiput(30,0)(0,-2.5){11}{\circle*{1}}
\end{picture} }
\put(50,30){$+$} \put(80,30){\begin{picture}(0,0)
\put(0,0){\circle*{5}} \put(60,0){\circle{5}}
\qbezier(0,3)(0,25)(30,25) \qbezier(60,3)(60,25)(30,25)
\qbezier(0,-3)(0,-25)(30,-25) \qbezier(60,-3)(60,-25)(30,-25)
\put(30,-35){$\bar{G}^r$} \put(30,30){$\bar{G}^a$}
\put(-15,0){$J^z_y$} \put(65,0){$j_x$}
\multiput(18,0)(0,2.5){10}{\circle*{1}}
\multiput(18,0)(0,-2.5){10}{\circle*{1}}
\multiput(42,0)(0,2.5){10}{\circle*{1}}
\multiput(42,0)(0,-2.5){10}{\circle*{1}}
\end{picture} }
\put(160,30){$+\cdots$}
\end{picture}
\caption{\label{fig:SHC} The spin Hall conductivity is calculated
in terms of diagrams. The first diagram contributes to
$\sigma_{sH}^0$, while the others contribute to the ladder
correction for spin Hall conductivity. $\bar{G}^r$ and $\bar{G}^a$
denote the average Green's functions taking over the distribution
of impurities. }
\end{figure*}
\begin{figure*}
\begin{picture}(0,56)(0,0)
\put(-200,10){ \begin{picture}(0,0) \put(-20,0){\vector(1,0){40}}
\put(-5,23){\line(5,2){10}} \put(5,23){\line(-5,2){10}}
\put(-3,30){M} \multiput(-10,0)(1,2.5){11}{\circle*{1}}
\multiput(10,0)(-1,2.5){11}{\circle*{1}}
\end{picture} }
\put(-170,25){$=$} \put(-155,25){$\displaystyle\int \displaystyle
\frac{1}{4\pi}d\theta d\phi \sin\theta \left(
\begin{array}{cc} \gamma\cos\theta & \sin\theta e^{-i\phi} \\
\sin\theta e^{i\phi} & -\gamma\cos\theta
\end{array}\right)\times$} \put(30,15){ \begin{picture}(0,0)
\put(-20,0){\vector(1,0){40}} \put(-5,23){\line(5,2){10}}
\put(5,23){\line(-5,2){10}}
\multiput(-10,0)(1,2.5){11}{\circle*{1}}
\multiput(10,0)(-1,2.5){11}{\circle*{1}}
\end{picture} }
\put(60,25){$\times\left(
\begin{array}{cc} \gamma\cos\theta & \sin\theta e^{-i\phi} \\
\sin\theta e^{i\phi} & -\gamma\cos\theta \end{array}\right)$}
\end{picture}
\caption{\label{fig:impurity}The Feynman diagram for the magnetic
impurity (left hand side of the diagram equation) is different from that for the
nonmagnetic impurity (right hand side). Each dot line
connected to the same impurity gives an extra matrix and the
integration $\int d\theta d\phi\sin\theta/4\pi $ for $\theta$ and
$\phi$ is performed.}
\end{figure*}

\subsection{Average Green's functions}

Now we calculate the average Green's functions in the
self-consistent Born approximation.
For simplicity, the Fermi energy $\mu$ is set to be the zero point of energy
since it can be combined in the dispersion relation.
Then the bottom of conduction band goes to negative infinite in
the large Fermi circle limit.
The free Green's function in chiral bases can be expressed as
\begin{eqnarray}\nonumber
G_{0(ch)}^r (\textbf{p},\omega) = \left( \begin{array}{cc}
\displaystyle \frac{1}{\omega-\varepsilon_+(p)+i\eta} & 0 \\
0 & \displaystyle \frac{1}{\omega-\varepsilon_-(p)+i\eta}
\end{array} \right), \\
\end{eqnarray}
where $\varepsilon_\pm(p)=\varepsilon(p)\mp p\alpha(p)$. The
Feynman diagram for the magnetic impurities is shown in Fig.~\ref{fig:impurity}.
The self-consistent Born equation for the averaged retarded Green's function
in chiral bases is given by
\begin{widetext}
\begin{eqnarray}\label{ret}
\bar{G}^r_{(ch)}(\textbf{p},\omega)
 = G_{0(ch)}^r (\textbf{p},\omega)
  & + &\sum_{\textbf{q}}\int d\theta d\phi
    \frac{\sin\theta}{4\pi} \frac{N u^2}{V^2}
      G_{0(ch)}^r (\textbf{p},\omega)
       U^\dagger (\textbf{p})
       \left(\begin{array}{cc} \gamma\cos \theta & \sin\theta e^{-i\phi}\\
                           \sin\theta e^{i\phi} &  -\gamma\cos\theta
       \end{array} \right)
        U(\textbf{q})
          \nonumber\\
 &\,& \bar{G}^r_{(ch)}(\textbf{q},\omega)
     U^\dagger (\textbf{q})
       \left(\begin{array}{cc} \gamma\cos \theta & \sin\theta e^{-i\phi}\\
                            \sin\theta e^{i\phi} &  -\gamma\cos\theta
       \end{array} \right)
         U(\textbf{p})
          \bar{G}^r_{(ch)}(\textbf{p},\omega),
\end{eqnarray}
\end{widetext}
where $N$ denotes the number of impurities. Eq.~(\ref{ret}) has a
solution
\begin{eqnarray}\nonumber
\bar{G}^r_{(ch)}(\textbf{p},\omega)=\left( \begin{array}{cc}
\displaystyle
\frac{1}{\omega-\varepsilon_+(p)+\frac{i}{2\tau}} & 0\\
0 & \displaystyle
\frac{1}{\omega-\varepsilon_-(p)+\frac{i}{2\tau}}
\end{array}
\right) , \\
\end{eqnarray}
where $\tau$ denotes the momentum-relaxation time and $1/\tau=n_i
u^2 2\pi N_F(\gamma^2+2)/3$, $n_i=N/V$ the impurity concentration
and $N_F$ the density of states at Fermi surface. Similarly, the
averaged advanced Green's function is
\begin{eqnarray}\nonumber
\bar{G}^a_{(ch)}(\textbf{p},\omega)=\left( \begin{array}{cc}
\displaystyle
\frac{1}{\omega-\varepsilon_+(p)-\frac{i}{2\tau}} & 0\\
0 & \displaystyle
\frac{1}{\omega-\varepsilon_-(p)-\frac{i}{2\tau}}
\end{array} \right). \\
\end{eqnarray}
The Green's functions in $\sigma_z$ bases are
\begin{eqnarray}\nonumber
\bar{G}^r(\textbf{p},\omega)=U(\textbf{p})\bar{G}^r_{ch}(\textbf{p},\omega)
U^\dag(\textbf{p}), \\[2mm]
\bar{G}^a(\textbf{p},\omega)=U(\textbf{p})\bar{G}^a_{ch}(\textbf{p},\omega)
U^\dag(\textbf{p}).
\end{eqnarray}
Then we calculate $\sigma^0_{sH}$, which corresponds to the
one-loop diagram in Fig. 2, at zero temperature and zero
frequency
\begin{widetext}
\begin{eqnarray}
 \sigma^0_{sH}= && \lim_{\omega\rightarrow 0} \frac{e}{V\omega}
  \int\frac{d\omega_1}{2\pi}\sum_{\textbf{p}}
    \textrm{Tr}\Bigl\{\theta(-\omega-\omega_1) J_y^z(\textbf{p})
    [\bar{G}^r(\textbf{p},\omega+\omega_1)-
     \bar{G}^a(\textbf{p},\omega+\omega_1)] j_x(\textbf{p})
        \bar{G}^a(\textbf{p},\omega_1)
           \nonumber\\
  && +~\theta(-\omega_1) J_y^z(\textbf{p}) \bar{G}^r(\textbf{p},\omega+\omega_1)
    j_x(\textbf{p})[\bar{G}^r(\textbf{p},\omega_1)- \bar{G}^a(\textbf{p},\omega_1)]
     \Bigr\}
      \nonumber\\
= && \frac{e}{8\pi}(1-\frac{1}{1+(\Delta \tau)^2}),
\label{SHC0}
\end{eqnarray}
\end{widetext}
where $\Delta=2p^{}_F \alpha(p_F)$ is the spin-orbit splitting at the
Fermi surface, and $p^{}_F\mid_{\varepsilon(p^{}_F)=\mu}$ is the Fermi
momentum.
In deriving the last line of Eq.~(\ref{SHC0}), we took
the integral with respect to $\omega_1$ and the summation with respect to $\mathbf{p}$
in the limit of large Fermi circle $\mu\gg 1/\tau, \Delta$
by means of the same method in Ref.~\cite{dimitrova}.
Clearly, $\sigma_{sH}^0$ is the same as that derived by Dimitrova
~\cite{dimitrova} for nonmagnetic impurities.
But it differs from $e/8\pi$ derived by
Inoue~\cite{Inoue06}, who takes the product of Green functions as
a $\delta$-function for simplicity. Such an approximation is valid
when $\Delta\tau\gg 1$, but it is inappropriate to take this
approximation before taking account of the vertex corrections to the
spin Hall conductivity.

\subsection{Vertex correction}

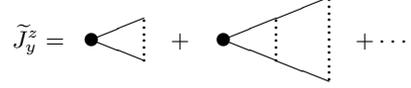
\begin{figure}[h]
\begin{picture}(0,70)(0,0)
\put(-80,20){$\widetilde{J}_y^z=$} \put(-50,23){\begin{picture}(0,0)
\put(0,0){\circle*{5}} \put(0,0){\line(5,2){20}}
\put(0,0){\line(5,-2){20}} \multiput(20,-8)(0,2.5){7}{\circle*{1}}
\end{picture} }
\put(-20,20){+} \put(0,23){\begin{picture}(0,0)
\put(0,0){\circle*{5}} \put(0,0){\line(5,2){40}}
\put(0,0){\line(5,-2){40}} \multiput(20,-8)(0,2.5){7}{\circle*{1}}
\multiput(40,-15)(0,2.5){13}{\circle*{1}}
\end{picture}}
\put(50,20){$+\cdots$}
\end{picture}
\caption{\label{fig:vertex}
The vertex of spin current with vertex corrections. }
\end{figure}

Furthermore, we calculate the vertex correction $\sigma^L_{sH}$ to the
spin Hall conductivity, which corresponds to the ladder diagram in
Fig.~2. Vertex corrections to the terms in Eq.(\ref{SHE}) with
two advanced or two retarded Green's functions vanish as
$\mu\gg1/\tau,\Delta$~\cite{dimitrova}. Only the vertex correction
with one advanced and one retarded Green's functions is
considered. The sum of ladder diagrams at $\omega=0$ gives
\begin{eqnarray}
\sigma^{L}_{sH}=\frac{-e}{2\pi V} \sum_{\textbf{p}}
\textrm{Tr}[\widetilde{J}^z_y
\bar{G}^r(\textbf{p},0)j_x(\textbf{p})\bar{G}^a(\textbf{p},0)],
\label{SHCL}
\end{eqnarray}
where the sum of the series of vertex corrections to the
$J_y^z(\textbf{p})$ is denoted by the matrix $\widetilde{J}_y^z$
(see Fig.~\ref{fig:vertex}).

\noindent
The self-consistent equation of $\widetilde{J}^z_y$ is
\begin{widetext}
\begin{eqnarray}
\widetilde{J}^z_y= \frac{n_i u^2}{V} \sum_{\textbf{p}} \int d\theta
d\phi \frac{1}{4\pi} \sin\theta
\left( \begin{array}{cc} \gamma\cos\theta & \sin\theta e^{-i\phi} \\
\sin\theta e^{i\phi} & -\gamma\cos\theta \end{array} \right)
\bar{G}^a (\textbf{p},0) (J^z_y(\textbf{p})+ \widetilde{J}^z_y )
\bar{G}^r(\textbf{p},0)
\left( \begin{array}{cc} \gamma\cos\theta & \sin\theta e^{-i\phi} \\
\sin\theta e^{i\phi} & -\gamma\cos\theta \end{array} \right).
\label{vertex1}
\end{eqnarray}
\end{widetext}
In the limit of large Fermi circle $\mu \gg \Delta,1/\tau$, the summation
over momentum in Eq.(\ref{vertex1}) can be evaluated by taking integral.
As a result, we have
\begin{eqnarray}\nonumber &&
(\widetilde{J}_y^z)_{\uparrow\uparrow}=(\frac{\gamma^2}{3}B+\frac{2}{3}C
) (\widetilde{J}_y^z)_{\uparrow\uparrow} +
(\frac{2}{3}B+\frac{\gamma^2}{3} C )
(\widetilde{J}_y^z)_{\downarrow\downarrow}, \\ \nonumber &&
(\widetilde{J}_y^z)_{\downarrow\downarrow}=(\frac{2}{3}B+\frac{\gamma^2}{3}
C ) (\widetilde{J}_y^z)_{\uparrow\uparrow} +
(\frac{\gamma^2}{3}B+\frac{2}{3}C)
(\widetilde{J}_y^z)_{\downarrow\downarrow}, \\ \nonumber &&
(\widetilde{J}_y^z)_{\uparrow\downarrow}=-\frac{\gamma^2}{3}A
-\frac{\gamma^2}{3}B(\widetilde{J}_y^z)_{\uparrow\downarrow}, \\
&& (\widetilde{J}_y^z)_{\downarrow\uparrow}=\frac{\gamma^2}{3}A
-\frac{\gamma^2}{3}B(\widetilde{J}_y^z)_{\downarrow\uparrow},
\end{eqnarray}
where $v^{}_F=d\varepsilon(p)/dp\mid _ {\varepsilon(p)=\mu}$ is the
Fermi velocity. A, B, and C are momentum integrals over products
of retarded and advanced Green functions.
\begin{eqnarray}
&& \nonumber A=\frac{-3iv^{}_F \Delta \tau}{4(\gamma^2+2)[1+(\Delta
\tau)^2]},
\\
&& \nonumber
B=\frac{3}{\gamma^2+2}(\frac{1}{2}+\frac{1}{2[1+(\Delta
\tau)^2]}),
\\
&& C=\frac{3}{\gamma^2+2}(\frac{1}{2}-\frac{1}{2[1+(\Delta
\tau)^2]}).
\end{eqnarray}
From these equations, we find
$(\widetilde{J}_y^z)_{\uparrow\uparrow}=(\widetilde{J}_y^z)_{\downarrow\downarrow}$
and
\begin{eqnarray}
(\widetilde{J}_y^z)_{\uparrow\downarrow}=-(\widetilde{J}_y^z)_{\downarrow\uparrow}
=\frac{ iv^{}_F \Delta \tau}{2+(8/\gamma^2+6)[1+(\Delta \tau)^2]}.
\end{eqnarray}
The vertex correction to the spin Hall conductivity is evaluated as
\begin{eqnarray}\nonumber
\sigma^{L}_{sH}&&=\frac{-ie\Delta\tau (\widetilde{J}^z_y)_{\uparrow\downarrow} }{4\pi v^{}_F}
[1-\frac{1}{1+(\Delta \tau)^2}]\\
&& \nonumber= \frac{(\Delta \tau)^2}{1+(4/\gamma^2+3)[1+(\Delta \tau)^2]} \sigma^0_{sH}. \\
\label{SHC1}
\end{eqnarray}
Summing Eq.(\ref{SHC0}) and Eq.(\ref{SHC1}),
we obtain the spin Hall conductivity
\begin{equation}
\sigma_{sH}=\sigma^{0}_{sH}+\sigma^{L}_{sH}= \frac{e}{8\pi}
\frac{(\Delta \tau)^2}{1+\displaystyle
\frac{3\gamma^2+4}{4\gamma^2+4}(\Delta \tau)^2} .\label{result}
\end{equation}
This is our main conclusion.

Now we discuss the relation between our work and a recent work on
such topics. Inoue et al.~\cite{Inoue06} considered the vertex
corrections due to magnetic impurities, which correspond to the
isotropic case $\gamma=1$ of our model. In this special case, our
result becomes
\begin{eqnarray}
\nonumber \sigma_{sH}&&= (1+\frac{(\Delta \tau)^2}{8+7(\Delta
\tau)^2})\sigma^0_{sH} \\ && =\frac{e}{8\pi} (1+\frac{(\Delta
\tau)^2}{8+7(\Delta \tau)^2}) (1-\frac{1}{1+(\Delta \tau)^2})
.\label{result1}
\end{eqnarray}
This differs from the result obtained by Inoue et al.~\cite{Inoue06}.
In their paper, the product of Green functions in Eq.(\ref{SHC0})
was taken to be a $\delta$-function so that their
$\sigma_{sH}^0$ and $\sigma_{sH}^L$ lack the factor
$(1-\displaystyle \frac{1}{1+(\Delta \tau)^2})$.
Such an approximation does not affect the total spin Hall conductivity
in the existence of nonmagnetic impurities
because the $\sigma_{sH}^0$ and $\sigma_{sH}^L$ cancel each other.
However, the $\sigma_{sH}^0$ and $\sigma_{sH}^L$ do not cancel each other
in the magnetically disordered system, and hence
the factor $(1-\displaystyle \frac{1}{1+(\Delta \tau)^2})$ enters the final result.
Even in the situation of $\tau \Delta \gg 1$,
the approximation condition of Ref.~\cite{Inoue06},
our $\sigma_{sH}|^{}_{\tau \Delta \gg 1}$
does not have the same lower order approximation as that
$\displaystyle\frac{e}{8\pi} (1+\frac{(\Delta \tau)^2}{8+7(\Delta
\tau)^2})$ in Ref.~\cite{Inoue06}.

\section{Summary and discussion}\label{sec:discussion}

Using Kubo's linear-response theory, we have evaluated, in the
ladder approximation, the vertex corrections of magnetic
impurities on the spin Hall conductivity in a Rashba-split 2DEG.
It was assumed that the magnetic impurities were short-ranged and
the orientations of their local moments were distributed
uniformly. Unlike the case of nonmagnetic impurities, the vertex
correction $\sigma^L_{sH}$ does not cancel $\sigma^0_{sH}$,
leading to a non-vanishing spin Hall conductivity (see
Eq.(\ref{result})) which depends upon the spin-orbit splitting
$\Delta$ and the momentum relaxation time $\tau$ in the limit of
large Fermi circle $\mu\gg\Delta,1/\tau$. The scattering changes
both the momentum and spin directions from a magnetic impurity and
lead to some correlation between them. Therefore, the average spin
tilting cannot cancel the spin current completely. In the dirty
limit $\tau\ll 1/\Delta$, $\sigma_{sH}$ goes to zero. While in the
clean limit $\tau\rightarrow \infty$, the spin Hall conductivity
is
\begin{eqnarray}
\lim_{\tau\rightarrow \infty}
\sigma_{sH}=\frac{e}{8\pi}\frac{4\gamma^2+4}{3\gamma^2+4},
\end{eqnarray}
which depends on the coefficient $\gamma$.
When $\gamma$ changes from $0$ to $\infty$, the clean limit of
$\sigma_{sH}$ changes from $e/8\pi$ to $e/6\pi$, and is $e/7\pi$
for the isotropic interaction.
This result differs from the universal value $e/8\pi$ for the ideal clean 2DEG
and also differs from the result $\sigma_{sH}=0$ in nonmagnetically
disordered system.

Concerning to the vertex corrections, the spin Hall conductivity
in nonmagnetically disordered or magnetically disordered systems does not
go back to the universal value $e/8\pi$ even if the impurity
concentration is arbitrarily small. This result comes from the
infinite size of system and infinite decoherence length
$L_\varphi$ assumed in calculation. As $\tau \rightarrow \infty$,
the corrected vertex $(\widetilde{J}^z_y)_{\uparrow\downarrow}$ goes
to zero, but the summation of Green's functions diverges because
of lack of a cut-off in momentum space. This divergence leads to a
non-vanishing $\sigma_{sH}^L$ and changes the total spin Hall
conductivity. In real systems, the quantum interference
contributing to the vertex corrections only happens inside the
decoherence length. So the spin Hall conductivity depends on the
decoherence time $\tau_\varphi$ when the momentum-relaxation time
is large enough to be comparable with it $\tau\sim\tau_\varphi$.
The $\sigma_{sH}$ derived in this work is only valid when $\tau$ is
not too large. Further studies are in progress.

\begin{acknowledgments}
This work was supported by the NSFC No. 10225419, 10674116, 10674117, 60471052; and
the Zhejiang Provincial Natural Foundation M603193.
\end{acknowledgments}

\end{document}